\documentclass[10pt,conference]{IEEEtran}
\IEEEoverridecommandlockouts
\usepackage{cite}
\usepackage{amsmath,amssymb,amsfonts}
\usepackage{graphicx}
\usepackage{textcomp}
\usepackage[table]{xcolor}

\usepackage{tikz}
\usetikzlibrary{decorations.pathreplacing,angles,quotes,shapes,shapes.geometric}

\usepackage{algpseudocode}
\usepackage{amsmath,amssymb,amsfonts}
\usepackage{url}
\usepackage{hyperref}
\usepackage{booktabs}
\usepackage{listings}

\lstdefinestyle{interfaces}{
  float=tp,
  floatplacement=tbp,
  abovecaptionskip=-5pt
}

\usepackage[ruled,vlined]{algorithm2e}

\SetKwInput{kwFailure}{Failure}
\usepackage{tabularx}
\usepackage{threeparttable} 
\usepackage{multirow} 
\usepackage{rotating} 
\usepackage{bigstrut} 
\usepackage{pgfplots}

\usetikzlibrary{arrows.meta}

\pgfplotsset{select coords between index/.style 2 args={
    x filter/.code={
        \ifnum\coordindex<#1\fi
        \ifnum\coordindex>#2\fi
    }
}}

\usepackage{soul}
\usepackage{todonotes} 

\newboolean{showcomments}
\setboolean{showcomments}{true}         

\ifthenelse{\boolean{showcomments}}
{\newcommand{\nb}[2]{
		\fbox{\bfseries\sffamily\scriptsize#1}
		{\sf\small:\textit{\textcolor{red}{#2}}}
	}
}
{\newcommand{\nb}[2]{}
	
}

\newcommand{\anonym}[1]{Anonymous for blind review}

\def\BibTeX{{\rm B\kern-.05em{\sc i\kern-.025em b}\kern-.08em
    T\kern-.1667em\lower.7ex\hbox{E}\kern-.125emX}}

\begin{document}

\title{Turning Federated Learning Systems Into Covert Channels}

\author{
	\IEEEauthorblockN{Gabriele Costa\IEEEauthorrefmark{1},
		Fabio Pinelli\IEEEauthorrefmark{1},	
		Simone Soderi\IEEEauthorrefmark{1},	 
		Gabriele Tolomei\IEEEauthorrefmark{2} \\}
	\IEEEauthorblockA{{\IEEEauthorrefmark{1}{IMT School for Advanced Studies, Lucca.} {\tt {firstname.lastname}@imtlucca.it}}\\
	{\IEEEauthorrefmark{2}{Sapienza University of Rome} {\tt lastname@di.uniroma1.it}}
	}
}

\maketitle

\newcommand{\tuple}[1]{(#1)}
\newcommand{\R}{\mathbb{R}}
\newcommand{\inst}{\boldsymbol{x}}
\newcommand{\X}{{\bf X}}
\newcommand{\dataset}{\mathcal{D}}
\newcommand{\features}{\mathcal{X}}
\newcommand{\labels}{\mathcal{Y}}
\newcommand{\params}{\boldsymbol{\theta}}
\newcommand{\model}{m^*}
\newcommand{\loss}{\mathcal{L}}
\newcommand{\trans}{\psi} 

\newcommand{\gabri}[1]{\todo[inline,color=red!60]{GT: #1}}

\begin{abstract}

Federated learning (FL) goes beyond traditional, centralized machine learning by distributing model training among a large collection of edge clients. 
These clients cooperatively train a global, e.g., cloud-hosted, model without disclosing their local, private training data.
The global model is then shared among all the participants which use it for local predictions.

In this paper, we put forward a novel attacker model aiming at turning FL systems into covert channels to implement a stealth communication infrastructure.
The main intuition is that, during federated training, a malicious sender can poison the global model by submitting purposely crafted examples.
Although the effect of the model poisoning is negligible to other participants, and does not alter the overall model performance, it can be observed by a malicious receiver and used to transmit a single bit.
\end{abstract}


\medskip
\hrule
\smallskip

Please cite this version of the paper: 

\noindent
G. Costa, F. Pinelli, S. Soderi and G. Tolomei ``Turning Federated Learning Systems Into Covert Channels''
in IEEE Access, doi: 10.1109/ACCESS.2022.3229124.
You may use the following bibtex entry:
\begin{verbatim}
@ARTICLE{CPST22covertchannel,
  author={Costa, Gabriele and Pinelli, 
  Fabio and Soderi, Simone and Tolomei, 
  Gabriele},
  journal={IEEE Access}, 
  title={Turning Federated Learning 
  Systems Into Covert Channels}, 
  year={2022},
  volume={10},
  number={},
  pages={130642-130656},
  doi={10.1109/ACCESS.2022.3229124}}
\end{verbatim}

\smallskip
\hrule
\smallskip
\section{Introduction}
\label{sec:intro}

\emph{Federated learning} (FL)~\cite{mcmahan2017googleai, mcmahan2017aistats} recently emerged as the leading technology for implementing distributed, large scale and efficient machine learning (ML) infrastructures. 
The main idea is that multiple clients connect to the FL system, and collaboratively train a shared, global model.
A frequent setting in FL is to have a centralized, e.g., cloud-hosted, server and many edge clients.
Clients and server iterate the execution of FL rounds, which consist of the following steps.
\begin{enumerate}
    \item The server sends the current, global model to every client and selects some of them for the next training round.
    \item Each selected client trains its local model using its own private data starting from the received global model, and sends the resulting local model back to the server.
    \item The server applies an \emph{aggregation function} to the local models received by clients and updates the global model.
\end{enumerate} 

FL allows clients to concurrently train a shared global model, without disclosing private training data.
Hence, FL provides great advantages in terms of both scalability and privacy.
Since this process smoothly integrates with ubiquitous, distributed infrastructures, it has been widely applied to IoT~\cite{Wu20personalized}, Fog computing~\cite{Zhou20privacy}, autonomous vehicles~\cite{Pokhrel20decentralized}, smartphones~\cite{Yang18google}, and wearable devices~\cite{Chen20fedhealth}.
As a result, nowadays billions of devices are connected to one or more FL systems.

The growing adoption of FL also raises security concerns, in particular, about the confidentiality, integrity, and availability of FL systems.
Indeed, several authors considered attack scenarios targeting FL systems.
For instance, in the last years some works investigated \emph{data poisoning} attacks, where an adversary pollutes the training set with maliciously crafted examples~\cite{jagielski2018sp}.
A similar scenario is that of \emph{model poisoning}, in which the attacker directly attempts to tamper with the global model parameters~\cite{Bhagoji19analyzing}.
The aim of both these types of attacks is usually to detriment the prediction accuracy of the learned model.
Also, a large body of work deals with privacy leakage that may expose the local data of some clients~\cite{Melis19exploiting}.

In this paper, we discuss a new attack scenario: we consider an adversary implementing a \emph{covert channel}~\cite{1973:lampson_covert_ch} over an FL system.
Covert channels allow an attacker to establish an illicit communication between two agents (e.g., two devices) that should stay isolated.
In theory, since no trust relationship exists among the clients, FL should not be intended to support the creation of covert channels. 
In practice, being shared among the participants, the global model can be turned into a communication channel. 
More specifically, two FL clients, i.e., a sender and a receiver, can agree on an aimed poisoning strategy that allows them to transfer one bit.

We start by describing our attacker model and the covert channel implementation strategy.
Our attacker only requires limited capabilities and, thus, it appears very realistic. 
As a matter of fact, communications are established by poisoning the training set of a single client, i.e., the sender.
Poisoned examples are crafted by modifying benign input examples through a simple, effective and efficient heuristics.

We explain the implementation strategy behind this attack  by considering an FL system, where clients collaboratively train a global model to recognize handwritten digits of the popular MNIST dataset~\cite{minstdataset}.
In our FL system, private training data is represented by a subset of the MNIST dataset that is randomly assigned to each client.
Such an implementation is often given as a template of a generic FL system in official tutorials.\footnote{For example, see \url{https://www.tensorflow.org/federated/tutorials/federated_learning_for_image_classification}.}



The main contributions of this paper are listed below.
\begin{itemize}
    \item A novel attacker model for FL-based covert channel.
    \item An implementation strategy working on any FL system. 
    \item A discussion of the main channel features.
\end{itemize}

The rest of the paper is organized as follows. 
Section~\ref{sec:background} describes the main background concepts used in this work.
In Section~\ref{sec:related}, we revise the literature about FL security and application level covert channels.
Section~\ref{sec:attacker-model} introduces our attacker model, and Section~\ref{SEC:IMPL} details the implementation of the covert channel.
In Section~\ref{SEC:CHARAT}, we describe the properties of our covert channel.
Finally, Section~\ref{sec:conclusion} concludes the paper.

\section{Background}
\label{sec:background}
In this section, we provide the reader with the essential context needed to understand the subject of this work.

\subsection{Machine \& Federated Learning}
We consider the {\em supervised learning} task as the reference example of a typical ML problem.
The goal of supervised learning is to estimate a function that maps an input to an output, based on a sample of observed input-output pairs, called \emph{examples}, which is usually referred to as \emph{training set}.

More formally, let $\dataset = \{(\inst_i, y_i)\}_{i=1}^n$ be a training set of $n$ examples. Each $\inst_i \in \features\subseteq \mathbb{R}^d$ is a $d$-dimensional vector of \emph{features} representing the $i$-th input and $y_i \in \labels$ is its corresponding output value.
Here, we focus on \emph{classification} problems where $\labels = \{1,\ldots,\ell\}$ and each $y_i$ is known as the \emph{class label} (as opposed to regression problems where $\labels \subseteq \R$).

Supervised learning assumes the existence of an unknown target function $g: \features \mapsto \labels$ that maps any feature vector to its corresponding output. 
The goal is therefore to estimate a function $\model$, namely a parametric model, that best approximates $g$ on $\dataset$.
More specifically, the optimal parametric model $\model$ is the one that minimizes the value of a loss function $\loss$, which measures the cost of replacing the true $g$ with $\model$ on the training set. 
In other words, learning $\model$ reduces to the following optimization problem, also known as empirical risk minimization (ERM)~\cite{vapnik1991nips}.
\begin{equation}
    \label{eq:erm}
    \model = \text{argmin}_{m}\loss(m, \dataset)
\end{equation}
Depending on the supervised learning task, different loss functions can be adopted. 
For example, cross-entropy is a commonly used loss function for classification~\cite{murphy2012ml}, whilst mean squared error is typically employed in regression settings~\cite{hastie2001sl}.

The standard framework above assumes that the actual training procedure, i.e., the optimizer used to solve \eqref{eq:erm}, runs on a centralized location where the whole dataset $\dataset$ is stored.
In the case of FL, instead, the learning process is distributed among several clients that collaboratively train a shared, global model with a centralized server acting as orchestrator.
Thus, the FL framework consists of a centralized server $S$ and a set of distributed, federated clients $\mathcal{C}$, such that $|\mathcal{C}|=n_c$. 
Each client $c\in \mathcal{C}$ has access to its own private training set $\mathcal{D}_c$, namely the set of its local labeled examples. 


The generic $t$-th round of FL comprises the following steps.
\begin{enumerate}
    \item $S$ sends the current, global model $m^{(t)}$ to every client and selects a subset ${\mathcal{C}}^{(t)}\subseteq \mathcal{C}$, such that $1 \leq |{\mathcal{C}}^{(t)}| \leq n_c$.
    \item Each selected client $c\in {\mathcal{C}}^{(t)}$ trains its local model $m_c^{(t)}$ by optimizing the same objective of \eqref{eq:erm} on its own private data $\dataset_c$, starting from $m^{(t)}$; the resulting $m_c^{(t)}$ is thus sent back to $S$.
    \item $S$ computes $m^{(t+1)} = \phi(\{m_c^{(t)}~|~c\in \mathcal{C}^{(t)}\})$ as the updated global model, where $\phi$ is an \emph{aggregation function} (e.g., FedAvg~\cite{mcmahan2017aistats} or one of its variants~\cite{lu2020spml}).
\end{enumerate} 

At the beginning, $m^{(0)}$ may be randomly initialized.
Then, FL rounds as the one described above are iteratively executed until convergence of the global model, i.e., until $t = T$ such that $m^{(T)} = \model$.
In practice, though, many FL models are continuously trained due to the highly dynamic nature of the infrastructure (e.g., new clients joining or leaving the system and fresh local data generated over time).

To simplify the notation, in the following we refer to $m$ as the global model and to $m_c$ as the local model of client $c$.
Furthermore, we call $m(\inst) = \hat{y}\in \labels$ (resp., $m_c(\inst) = \hat{y}_c\in \labels$) the global (resp., local) model prediction on input $\inst$.

\subsection{Channels and communication quality}
\label{sec:bkg-covert}

\begin{figure}[t]
    \centering
    \includegraphics[width=\columnwidth]{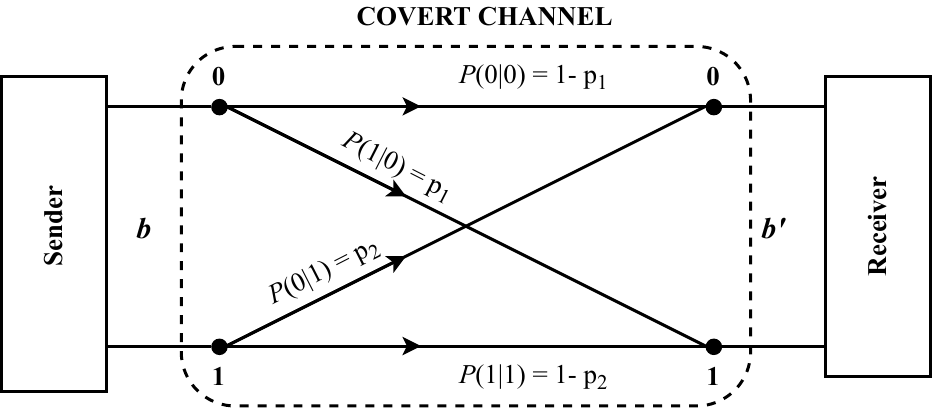}
    \caption{Binary memoryless covert channel.}
    \label{fig:bsc}
\end{figure}

In this paper, we are interested in \emph{binary memoryless channels} (BMC)~\cite{Proakis2007digital}.
Briefly, an BMC has discrete input $b$ and output $b'$ such that $b, b' \in \{0,1\}$.
We describe the relationship between channel's input and output through the conditional probability $P(\mathbf{b}|\mathbf{b})$, where $\mathbf{b} = \{0,1\}$.
Figure~\ref{fig:bsc} depicts a generic BMC in which $P(1|0) = p_1$ and  $P(0|1) = p_2$ are the probabilities for input/output bit inversion errors. 
Consequently, their complements give the probability of receiving the correct bit, e.g., $P(0|0) = 1 - p_1$. 
For a BMC, the \emph{channel capacity} $C$ is the maximum communication rate that sender and receiver can reach over the channel. 
Following~\cite{Proakis2007digital}, $C$ is computed as follows.
%
%
%
\begin{equation}
    C = \max_{P} \mathbb{I}(x;y) =\sum\limits_{\substack{x\in \mathbf{b} \\ y\in \mathbf{b}}} P(x) P(y|x) \log \Bigg(\frac{P(y|x)}{P(y)}\Bigg), \label{EQ:CH_C}
\end{equation}
where $\mathbb{I}(x;y)$ is the mutual information between the input $x$ and output $y$. 
%
%
From~\eqref{EQ:CH_C}, we obtain the channel capacity as
%
%
%
\begin{align}
\label{EQ:COVERCH_C}
    C = &\: 1 + \sum_{i = 1,2} \frac{1}{2} (p_i + (1-p_i) \log_2(1-p_i)) = \nonumber \\
      = &\: 1 - \sum_{i = 1,2} \frac{ H(p_i)}{2} =  1 - \frac{ H(p_1)}{2} - \frac{ H(p_2)}{2},
\end{align}
where $H$ is the binary Shannon entropy function.


Communication quality is typically measured in terms of \emph{bit error rate} (BER)
and \emph{signal-to-noise ratio} (SNR).
In general, BER is obtained as the number of bit inversions, e.g., due to channel noise, over the total number of transmitted bits.
Instead, SNR is defined as the ratio of signal power to the noise power, i.e., $s^2/n^2$, where $s$ is the channel signal and $n$ is the noise. 
In general, when signal and noise are modeled by means of two random variables, called $S$ and $N$ (respectively), we have that $s^2 = E[S^2]$ and $n^2 = E[N^2]$, where $E[\cdot]$ denotes the expected value.
Moreover, when $N$ has zero mean, i.e., $E[N] = 0$, $n^2$ reduces to $\sigma^2_n = Var(N) = E[N^2] - E[N]^2$, that is $E[N^2]$ is equal to the variance of $N$.

\section{Related work}
\label{sec:related}


Our proposal consists of an application level covert channel implemented through an adversarial attack to an FL system.
Although some authors reviewed the security challenges of FL, e.g., see~\cite{Kairouz21advances} for an extensive discussion, we are not aware of papers describing similar, covert channel scenarios.
Thus, in the following, we revise some related work about \emph{adversarial attacks to ML}, and \emph{application level covert channels} which are closer to our proposal. 
%

\subsection{Adversarial attacks to ML}
\label{subsec:adversarial-attacks}

There exists a large body of work investigating the security of both traditional, i.e., centralized, and federated ML against so-called {\em adversarial attacks}~\cite{dalvi2004kdd,huang2011aisec,biggio2018pr,nasr2019sp,pitropakis2019csr,lyu2020arxiv,jere2020sp,tolpegin2020arxiv,goldblum2020arxiv}.
Adversarial attacks can have different targets.
For instance, {\em model inversion attacks}~\cite{fredrikson2014usenix,fredrikson2015ccs}, {\em membership inference attacks}~\cite{shokri2017sp,pyrgelis2018ndss,melis2019sp}, and {\em property inference attacks}~\cite{ateniese2015ijsn, ganju2018ccs} aim to violate the {\em confidentiality} of user's private training/test data.
Also, it has been reported that an attacker can compromise the confidentiality/intellectual property of a model provider by stealing its model parameters and hyperparameters~\cite{bin2016www, tramer2016usenix, binghui2018sp}.

From this perspective, our attacker model belongs to adversarial attacks that tamper with the integrity and the performance of a predictive model~\cite{barreno2006asiaccs}.
These attacks are classified according to the stage(s) of the ML pipeline in which they occur. 
In particular, attacks can happen at training time only, both at training and at test time, or at test time only. 
Those are called \emph{poisoning}, \emph{backdoor}, and {\em evasion} attacks, respectively.

From another viewpoint, depending on the attacker's goal, adversarial attacks can be further classified as {\em untargeted} (or random)~\cite{rubinstein2009imc,biggio2012icml,xiao2015icml,li2016nips,yang2017ndss,jagielski2018sp} or {\em targeted}~\cite{nelson2008usenix,shafahi2018nips,gu2019ieee,aghakhani2020arxiv}.
The former aims to reduce the overall accuracy of the learned model at inference time, regardless of what specific testing examples get incorrectly classified. 
The latter forces the learned model to output attacker-desired labels for certain testing examples, e.g., predicting spam messages as non-spam, while not altering the output for other examples.
Since targeted attacks have to do with a specific goal, they usually require rather strong attacker's capabilities.

In the following we provide a detailed overview of the most prominent adversarial attacks.

\paragraph{Poisoning Attacks}
\label{par:poisoning-attacks}
To compromise the performance of a predictive model, poisoning attacks can target two components of the training stage, i.e., the training dataset and the learning process.
The former are known as {\em data poisoning} attacks~\cite{nelson2008usenix,huang2011aisec,biggio2012icml,fang2018acsac,jagielski2018sp}.
The latter are referred to as {\em model poisoning} attacks~\cite{bhagoji2019pmlr,fang2020usenix}.

\noindent
\textbf{Data Poisoning}. These attacks pollute the training dataset by injecting it with new malicious examples or by corrupting existing ones. 
There are two main types of data poisoning attacks, called \emph{(i)} \emph{clean-label}~\cite{shafahi2018nips} and \emph{(ii)} \emph{dirty-label}~\cite{gu2017arxiv}, respectively.
In clean-label poisoning attacks, the adversary has no control over the labeling process. 
The attacker simply injects a small number of slightly perturbed examples (whose labels remain correct) into the training set of the victim.
These attacks have been proven effective only when the attacker has complete knowledge of the victim's model, i.e., under \emph{white-box} assumption~\cite{pitropakis2019csr}.
Such a knowledge is needed to craft the malicious examples~\cite{suciu2018usenix}. 
More recently, clean-label poisoning attacks for unknown, i.e., \emph{black-box}~\cite{papernot2017asiaccs}, deep image classifiers have been also explored~\cite{zhu2019pmlr}.

In dirty-label poisoning the adversary can introduce a number of training instances to be misclassified with a specific label in a targeted way. 
An example of dirty-label poisoning is the {\em label-flipping} attack~\cite{biggio2012icml,fung2018arxiv}.
Here, the labels of honest training examples of one class are flipped to another class, while the features of the data are kept unchanged. 
For instance, a malicious agent can poison the training set of a handwritten digit recognition system by flipping all 1s to 7s. 

In the FL setting, it is common to assume that the attacker controls some clients and their training sets. 
Thus, dirty-label attacks have been more often considered. In~\cite{tolpegin2020arxiv}, the authors study targeted label flipping attacks on FL. 
They find that poisons injected late in the training process are significantly more effective than those injected early. 
Other proposals adopt a bi-level optimization approach for poisoning multi-task FL~\cite{sun2020arxiv} and GAN-generated poisons~\cite{zhang2019bdse}.

\noindent
\textbf{Model Poisoning}. Differently from data poisoning attacks, these attacks threaten the learning process directly, e.g., by changing some model parameters. 
Model poisoning is generally perceived as difficult to implement in centralized ML systems as it requires the adversary to access the target model, i.e., assuming either \emph{grey-box} or \emph{white-box} knowledge. 
On the other hand, model poisoning becomes rather feasible in the case of FL, where a malicious client has direct influence over the jointly-trained global model via its local parameters updates~\cite{mahloujifar2019pmlr, bhagoji2019pmlr,fang2020usenix}.
As with any poisoning attack, the adversary's goal is to cause wrong predictions of the FL model. 
However, she/he aims to force classification errors at inference time and without modifying test examples.
In this respect, it is opposed to backdoor and evasion attacks (which are discussed below).
In model poisoning, the misclassification results from the adversarial corruption of the training process, which can be achieved either by \emph{gradient} or {\em learning rule} manipulation. 
In gradient manipulation, the adversary poisons local model gradients (or model updates), which are then sent to the central server for aggregation, thereby jeopardizing the global model performance~\cite{blanchard2017nips}.
In learning rule manipulation, instead, the attacker corrupts the actual training logic. 
In some cases, these model poisoning attacks have proven more effective than data poisoning, and an attacker can compromise the global model even when controlling a single client.
For instance, in~\cite{bhagoji2019pmlr} the authors successfully achieve a stealthy targeted model poisoning attack by adding a penalty term to the objective function to minimize the distance between malicious and benign weight update distributions. 

\paragraph{Backdoor Attacks}
\label{par:backdoor-attacks}

Backdoor attacks  -- also known as trojan attacks -- exploit the adversary's capability of having (limited) access to input examples also at test time~\cite{goldblum2020arxiv}.
Hence, backdoor attacks exceed poisoning attacks, since the adversary can manipulate both training and test inputs.
Therefore, backdoor attacks are often considered more disruptive toward the victim model.


Like for standard poisoning attacks, we can distinguish between backdoor attacks affecting the data or the model. 
The former are referred to as backdoor data poisoning and consist of adding  attacker-chosen examples to the training set.
The attacker examples contain a particular {\em trigger}~\cite{chen2017arxiv,liu2018ndss}, i.e., a distinguished feature that activates the backdoor. 
The model learned on such poisoned training set will embed a backdoor, which the attacker exploits at test time by submitting examples that contain the same trigger.
Instead, backdoor model poisoning requires a stronger threat model, where the attacker can get direct access to the learning system and change the model's internals (i.e., parameters and architecture) to embed a backdoor~\cite{dumford2018arxiv,ji2018ccs}.

Interestingly enough, backdoor attacks have been proven ineffective under FL settings~\cite{Bagdasaryan20backdoor}.
The main obstacle is that aggregation involves many clients and, assuming that the attacker only controls a minority, the effect of the adversarial updates is weakened. 
To overcome this limitation, the authors of~\cite{Bagdasaryan20backdoor} consider a model replacement approach, where the attacker scales up a malicious model update to increase its effect on the aggregation function.

In~\cite{xie2020iclr}, the authors propose distributed backdoor attacks, which better exploit the decentralized nature of FL. 
Specifically, they decompose the backdoor pattern for the global model into multiple distributed small patterns, and inject them into training sets used by up to 40\% adversarial participants at each federated round.
Although more effective than global backdoor trigger injection, this approach comes at the price of controlling a significant subset of the FL clients.


\paragraph{Evasion Attacks}
\label{par:evasion-attacks}
Adversarial attacks that occur only at test time are called evasion attacks.
Here, the goal of the adversary is still to fool an ML model yet without tampering it at training time.
In fact, evasion attacks make use of so-called {\em adversarial examples}~\cite{barreno2006asiaccs,szegedy2014iclr}, i.e., carefully-crafted (minimal) perturbations of test instances that cause prediction errors (either targeted or untargeted) when input to a legitimately trained model.

Evasion attacks may look similar to backdoor poisoning attacks~\cite{biggio2013mlkdd}. 
However, the key difference between the two is that evasion attacks exploit the decision boundaries learned by an uncorrupted model to construct adversarial examples that are misclassified by the model. 
In contrast, backdoor attacks intentionally shift these
decision boundaries as a result of a jeopardized training process, so that certain examples get eventually misclassified~\cite{goldblum2020arxiv}.
 
Several works have explored evasion attacks in the context of computer vision~\cite{carlini2017sp}, where adversarial examples are obtained by adding random noise to test images.
Even though such images look legitimate to a human, they are wrongly classified by the image recognition system.
Also, more recent works investigate the applicability of evasion attacks to malware classification~\cite{suciu2019spw}.

In the FL setting, the global model maintained by the server suffers from the same evasion attacks as in the conventional ML setting when the target model is deployed as a service. Moreover, at each FL training round the global model sent to the federated clients is exposed as a white-box to any malicious participant. Thus, FL requires extra efforts to defend against white-box evasion attacks~\cite{lyu2020arxiv}.

In this work, we consider an attack scenario partially related to the one presented in~\cite{Bagdasaryan20backdoor}.
As a matter of fact, our covert channel is implemented by means of maliciously crafted examples that carry a trigger as in backdoor attacks.
In particular, our adversary $(i)$ crafts malicious examples carrying a trigger (see Section~\ref{sec:calibration}), and $(ii)$ poisons the federated model to transmit one bit through each trigger (see Section~\ref{SEC:BIASED_ID}).

\subsection{Application level covert channels}
\label{subsec:covert-channels}


In a general sense, a covert channel is any communication channel that is not intended for information transfer~\cite{1973:lampson_covert_ch}.
Although we are not aware of FL-based covert channels, some authors already investigated the implementation of covert channels at application level.
Most authors considered \emph{encapsulation} of hidden communications in application-layer network protocols.
Being the main application level protocol, HTTP is the primary target for covert channel implementations, e.g., see~\cite{Bauer03covert}. 
Nevertheless, the entire TCP/IP ecosystem can be at risk, and we refer the interest reader to~\cite{Mileva14covert} for a survey.
More recently, also web applications were proposed for the implementation of covert channels.
For instance, in~\cite{Selvi12covert} the author considers social networks such as Facebook and Twitter.
However, since this kind of covert channels rely on already existing communications between devices, in practice, they usually do not break any sandbox policy.\footnote{On the contrary, they are very relevant, for instance, when considering inter-process channels as in~\cite{Chandra2015android}.}
Also, as the authors of~\cite{Zander07covert} point out, most of these covert channels can be detected and some effective countermeasures exist, e.g., packet inspection can be used to detect illegal traffic.
Possibly for these reasons, application level covert channels appear to be less frequently considered in the literature.

Loosely speaking, also our proposal relies on a sort of encapsulation mechanism.
However, here we do not wrap information inside protocol messages. 
Rather, we embed information inside FL models, which prevents standard detection techniques based on traffic inspection.

\section{Attacker Model}
\label{sec:attacker-model}

In this section, we present our attacker model.
The goal of the attacker is to establish a covert channel between two clients, namely \emph{Sender} and \emph{Receiver}, of an FL infrastructure.
In terms of capabilities, our adversary resembles that of~\cite{Bagdasaryan20backdoor}.
Here, we assume that both Sender and Receiver are controlled by the attacker.
For instance, think of Sender as a malware-compromised device and Receiver as the malware command and control server.
Although they are compromised, we do not assume that the attacker can tamper with the standard FL infrastructure behavior, i.e., Sender and Receiver follow the FL client protocol. 
More precisely, Sender is only allowed to poison its local dataset and Receiver can only classify examples using its own local model.
In particular, we stress that here we assume the attacker to operate in \emph{black-box} mode, i.e., we assume that Sender and Receiver can neither inspect nor tamper with their local models.
It is worth noticing that these assumptions are very general and they apply to most FL systems. 


\begin{figure}[t]
    \centering
     \begin{tikzpicture}[block/.style={draw,rectangle,rounded corners=0.5mm,minimum width={1cm}, minimum height={1cm}},sblock/.style={draw,rectangle,rounded corners=0.5mm,minimum width={1cm}, text height=0.8cm, minimum height={1.2cm}}]
    
    \node[block] (Sender) at (0,0) {\includegraphics[width=1.2cm]{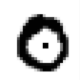}};
    \node[sblock] (Server) at (3.5,-2) {\small $m = \phi(\{m_s, m_1, m_2\})$};
    \node[block] (Receiver) at (7,0) {\includegraphics[width=1.2cm]{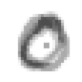}};
    
    \node[block] (Client1) at (0,-2) {\includegraphics[width=1.2cm]{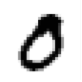}};
    \node[block] (Client2) at (0,-4) {\includegraphics[width=1.2cm]{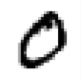}};
    
    \node[block] (Client3) at (7,-2) {\includegraphics[width=1.2cm]{pics/m.png}};
    \node[block] (Client4) at (7,-4) {\includegraphics[width=1.2cm]{pics/m.png}};
    
    \node (Attacker) at (3.5,0) {\includegraphics[width=1.2cm]{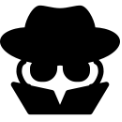}};
    
    \node at (0,-0.9) {Sender};
    \node at (0,-2.9) {Client 1};
    \node at (0,-4.9) {Client 2};
    
    \node at (3.5,-2.9) {FL Server};
    \node[cylinder, draw, shape aspect=.4,  
      cylinder uses custom fill, cylinder end fill=gray!50, 
      minimum height=2cm,
      cylinder body fill=gray!25, 
    scale=1] (Channel) at (3.4,-1.7) {covert channel};
    
    \node at (7,-0.9) {Receiver};
    \node at (7,-2.9) {Client 3};
    \node at (7,-4.9) {Client 4};
    
    \node at (3.5,-0.75) {Attacker};
    
    \draw[thick, -latex] (Sender) -- node[midway,below left=-2mm] {$m_s$} (Server.170);
    \draw[thick, -latex] (Client1) -- node[midway, below] {$m_1$} (Server.west);
    \draw[thick, -latex] (Client2) -- node[midway,below right] {$m_2$} (Server.190);
    
    \draw[thick, -latex] (Server.10) -- node[midway,below right=-1mm] {$m$} (Receiver);
    \draw[thick, -latex] (Server.east) -- node[midway, below] {$m$} (Client3);
    \draw[thick, -latex] (Server.350) -- node[midway,below left] {$m$} (Client4);
    
    \draw[thick, -latex] (Attacker) -- node[pos=0.4,below=-1mm] {\includegraphics[width=0.7cm]{pics/dotted0.png}} (Sender);
    \draw[thick, -latex] (Attacker) -- node[pos=0.4,below=-1mm] {\includegraphics[width=0.7cm]{pics/dotted0.png}} (Receiver);
    
    \node at (1.85,0.45) {\parbox{1cm}{\centering\small crafted \\ example}};
    \node at (5.05,0.45) {\parbox{1cm}{\centering\small crafted \\ example}};
    
    \path[thick,-latex]
    (Sender.340) edge[dashed,bend right=10] node [above] {$b$} (Channel.west);
    \path[thick,-latex]
    (Channel.east) edge[dashed,bend right=10] node [above] {$b$} (Receiver.200);
    
    \end{tikzpicture}
    \caption{Overview of the attacker model.}
    \label{fig:attacker-model}
\end{figure}


The overall attacker model is schematically depicted in Figure~\ref{fig:attacker-model}.
At each federated round, the FL server randomly selects a subset of clients.
Selected clients work as expected, i.e., they use their own private datasets to train their local models starting from the last global model received by the server (see Client 1 and Client 2 in Figure~\ref{fig:attacker-model}).
Then, FL clients upload their newly trained local models to the server, which aggregates them into an updated global model $m$ through the aggregation function $\phi$.
At the end of each federated round, \emph{all} the clients receive a copy of $m$. 
When selected, Sender (top left of Figure~\ref{fig:attacker-model}) poisons its local model $m_s$ by training it with some malicious, attacker-provided examples.
Its goal is to transmit a bit $b$ by inducing a perturbation of the global model that Receiver can test.
On the other hand, Receiver (top right of Figure~\ref{fig:attacker-model}) uses $m$ to classify test examples, e.g., the same malicious examples used by Sender.
According to the classification outcome, Receiver deduces whether 0 or 1 was sent.
Implementing such a covert channel is non-trivial and also depends on the underlying FL system.
We discuss the implementation details in the next section.

%
%
\section{Covert channel implementation}
\label{SEC:IMPL}

\begin{figure}[t]
  \centering
  
  \begin{tikzpicture}[every node/.append style={very thick,rounded corners=0.2mm}]
    
    \node[draw,rectangle] (Receiver) at (0,0) {Receiver};
    
    \node[draw,rectangle] (Server) at (3,0) {FL Server};
    
    \node[draw,rectangle] (Sender) at (6,-2) {Sender};
    
    \draw [very thick] (Receiver)--++(0,-5.2);
    \draw [very thick] (Server)--++(0,-5.2);
    \draw [very thick] (Sender)--++(0,-3.2);
    
    \draw [latex-,thick] (0,-1)--node [auto] {$m^{(1)}$}++(3,0);
    \node at (1.5,-1.2) {$\vdots$};
    \draw [latex-,thick] (0,-2.1)--node [auto] {$m^{(f)}$}++(3,0);
    
    \draw [-,thick] (0,-2.8)--node [auto] {$(f,\widetilde{\inst},h,l)$}++(2.9,0);
    \draw [-latex,thick] (3.1,-2.8)--node [auto] {}++(2.9,0);
    \draw[black, thick] (3.1,-2.8) arc (0:180:0.1);
    
    \draw [latex-,thick] (0,-3.5)--node [auto] {$m^{(f+1)}$}++(3,0);
    \node at (1.5,-3.7) {$\vdots$};
    \draw [latex-,thick] (0,-4.6)--node [auto] {$m^{(2f)}$}++(3,0);
    
    \draw [-latex,thick] (3,-3.8)--node [auto] {selected}++(3,0);
    \draw [latex-,thick] (3,-4.3)--node [auto] {$m_b$}++(3,0);
    
    \draw[-latex,thick] (6,-3.9) -- +(0.8,0) |- (6,-4.2);
    
    \draw[-latex,thick] (0,-4.7) -- +(-0.8,0) |- (0,-5);
    
    \node at (6.6,-3.65) {write $b$};
    \node at (-0.6,-4.45) {read $b$};
    
    \draw[decoration={brace,mirror,amplitude=5pt,raise=5pt},decorate,very thick] (-1,-0.8) -- node[left=5pt] {} (-1,-2.2);
    \node[label={[text depth=-1ex,rotate=90]right:calibration}] at (-1.8,-2.4) {};
    \draw[decoration={brace,mirror,amplitude=5pt,raise=5pt},decorate,very thick] (-1,-3.3) -- node[left=5pt] {} (-1,-5.2);
    \node[label={[text depth=-1ex,rotate=90]right:transmission}] at (-1.8,-5.3) {};
    \end{tikzpicture}

    \caption{Communication protocol phases.}
    \label{fig:msc}
\end{figure}
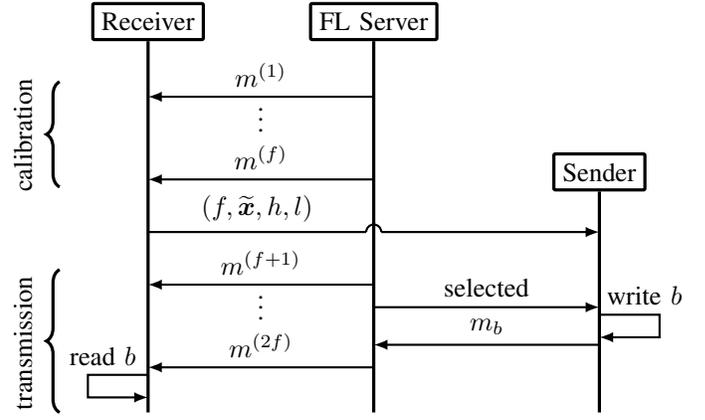


In this section, we detail the implementation strategy for creating the previously described covert channel.
Without loss of generality, every implementation is based on the abstract protocol schematically depicted in Figure~\ref{fig:msc}. 

A transmission starts with a \emph{calibration} phase during which Receiver observes the global model updates $m^{(1)}, \ldots, m^{(f)}$ (for $f$ FL rounds).
Eventually, Receiver computes \emph{channel parameters} (see Section~\ref{sec:calibration}) to be shared with Sender.
For instance, these parameters can be provided through a secondary channel or hard-coded in Sender before its deployment.
Then, Sender and Receiver synchronize on transmission frames of size $f$ to send a bit $b$.
During each frame, when selected, Sender trains its local model according to the channel parameters and the bit to be transmitted.
In the meanwhile, Receiver monitors the global model updates and, at the end of the frame, it tests the received bit.

Below, we discuss the implementation of both the calibration and transmission phases.

%
%
\subsection{Calibration of channel parameters}
\label{sec:calibration}
The first parameter to be determined is the size of the transmission frame $f$, i.e., the number of FL rounds necessary to transmit a single bit.
Assuming that the attacker ignores the internals of the FL server, i.e., current number of clients ($n_c$) and client selection probability ($p_c$), Receiver must estimate $f$.
Clearly, a small $f$ decreases the probability for Sender to be selected (which is necessary for the transmission to take place), while large $f$ reduces the transmission rate.
Under the assumption that clients are uniformly selected at random, Receiver takes advantage of its selection notifications to estimate $f$.
For instance, it can set $f$ as the number of rounds that it takes to be selected $T$ times (for a constant $T$).

Another channel parameter to be determined is the crafted example $\widetilde{\inst}$ carrying the trigger that Sender will use to poison its local training set.
To generate $\widetilde{\inst}$, Receiver applies a \emph{linear transformation function}\footnote{Although we do not explicitly prove it, the reader can easily check that all the transformation functions presented in the following are linear, since they reduce to finite sums of matrices.} $\trans$ to a subset of $k$ randomly selected examples from its local training set. 
Intuitively, $\widetilde{\inst} = \trans(\inst_1, \ldots, \inst_k, \alpha)$ means that, starting from $k > 0$ examples $\inst_1, \ldots, \inst_k$, $\trans$ returns a new example $\widetilde{\inst}$, where $\alpha \in [0,1]$ is a parameter that controls the transformation.

\begin{figure}[t]
    \centering
    \begin{tabular}{|c|c|c|}
    \toprule
    \includegraphics[width=0.28\columnwidth]{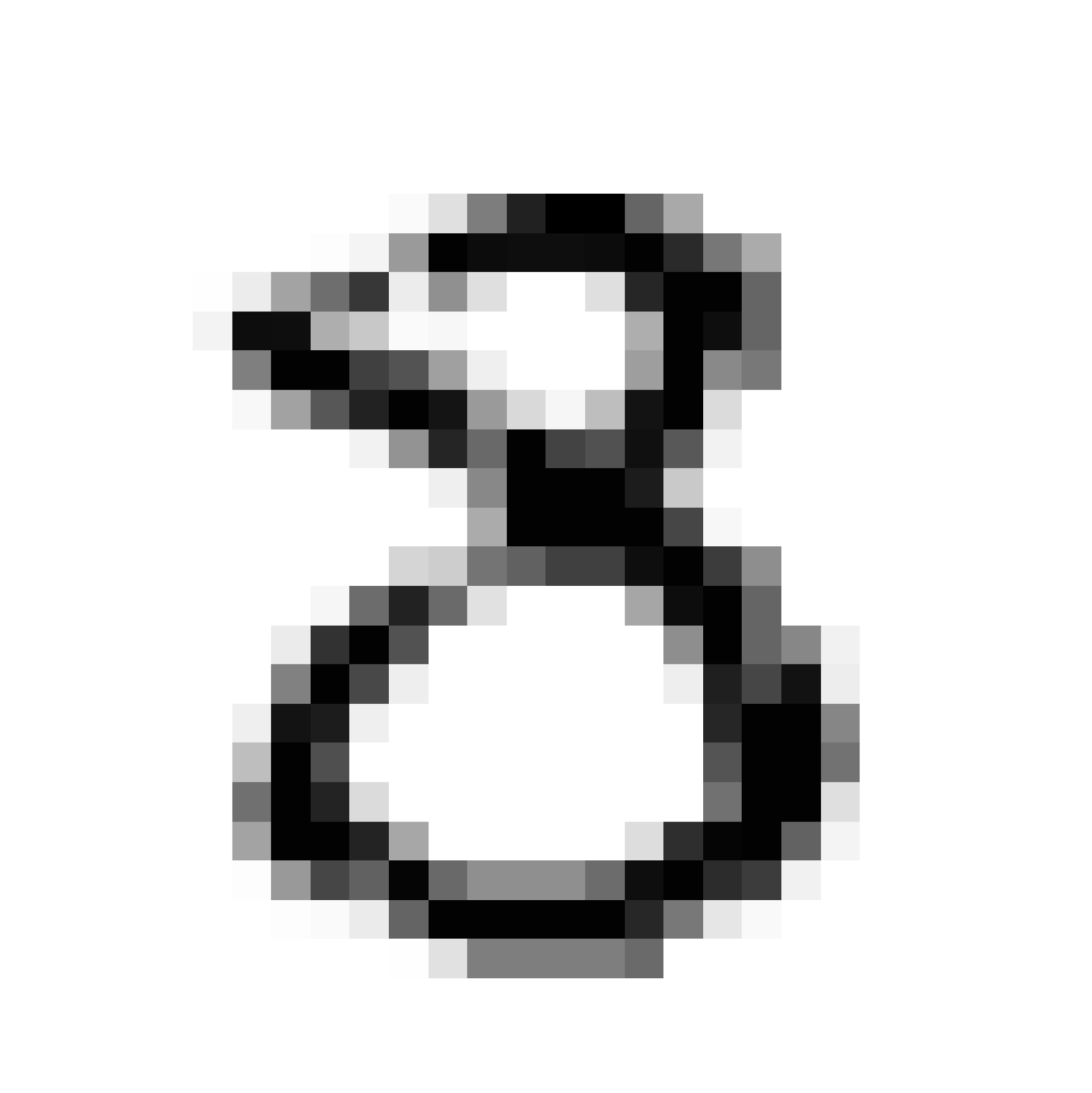}   &
    \includegraphics[width=0.28\columnwidth]{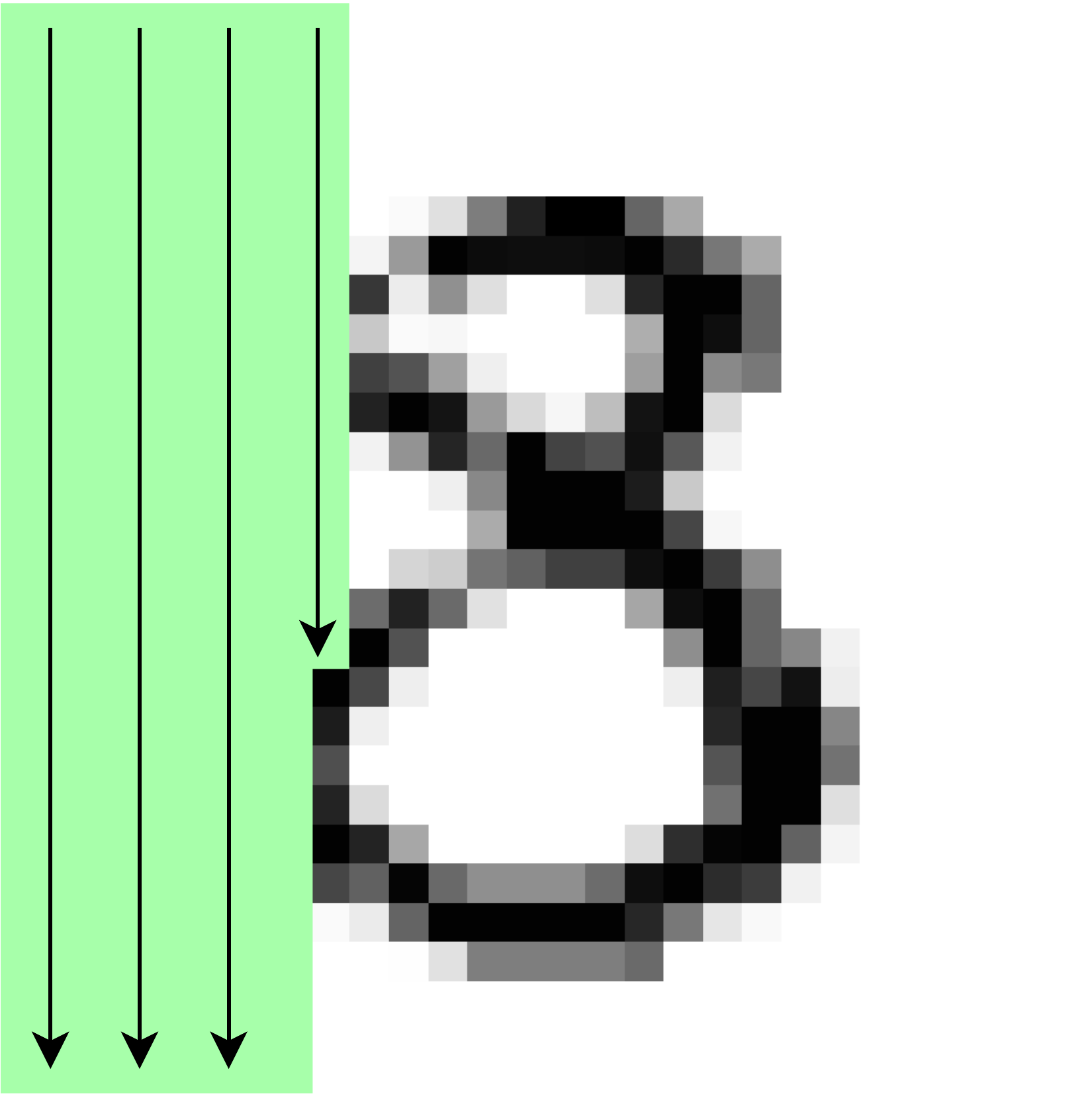} &
    \includegraphics[width=0.28\columnwidth]{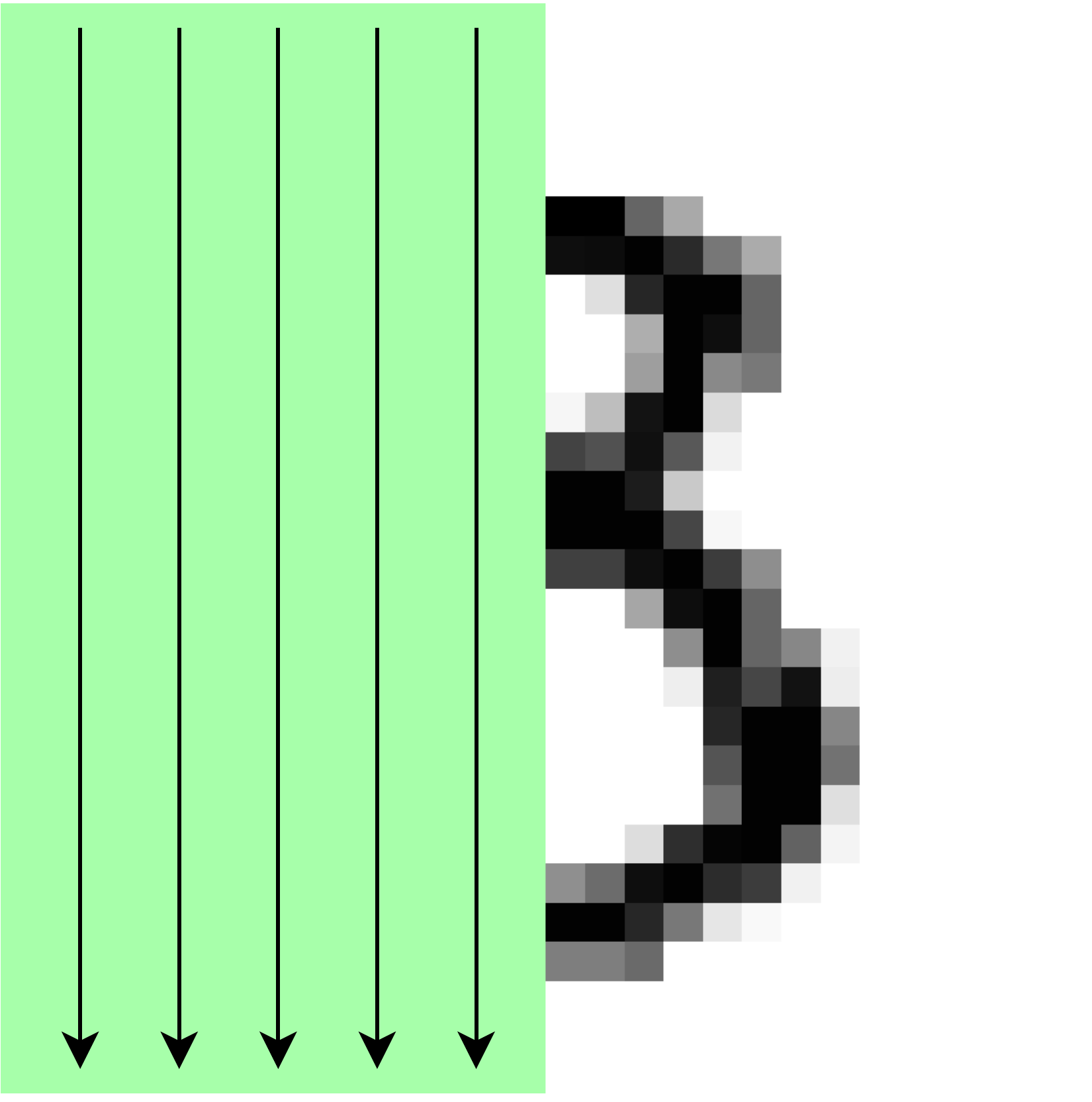} \\
    \bottomrule
    \end{tabular}
    \caption{Linear transformation $\psi_e$ of an example with $\alpha = 0.3$ (middle) and $\alpha = 0.5$ (rightmost).}
    \label{fig:sample-craft}
\end{figure}

To provide an intuition of this process, we put forward an example taken from the MNIST dataset.
Consider $\boldsymbol{x}$ to be the representation of a generic MNIST input image, i.e., a $28\!\times\!28 $ matrix of pixels flatten into a $784$-dimensional vector.
We define $\trans_e(\inst, \alpha)$ as the function erasing, from left to right, a fraction $\alpha$ of a single example image $\boldsymbol{x}$ (i.e., $k = 1$).
For instance, when $\alpha = 0.3$, $\trans_e$ sets to 0 the $235$ values associated with the leftmost pixels of the target image vector. 
More formally, $\psi_e(\inst,\alpha) = \inst - \inst^{(\alpha)}$ where $\inst^{(\alpha)} = [x^1, \ldots, x^{\lfloor\alpha \cdot 784 \rfloor}, 0, \ldots, 0]$ and $x^i$ is the $i$-th element of $\inst$.
The behavior of $\trans_e$ is shown in Figure~\ref{fig:sample-craft}, where we highlighted the erased portion of the original example.

Intuitively, the attacker can select a function $\trans$ which is based on some \emph{semantic} property of the classification domain.
For instance, in the previous example $\trans_e$ encodes the simple fact that the right half of an handwritten 8 looks like a 3.
Other similar functions can be found, e.g., in~\cite{castro2019morphomnist}.

Receiver repeatedly applies $\trans$ to the selected examples until it identifies an \emph{edge} example $\widetilde{\inst}$.
Formally, given Receiver's local model $m_r$ and an arbitrarily small $\varepsilon > 0$, we look for $\widetilde{\inst}=\trans(\inst_1, \ldots, \inst_k, \bar{\alpha})$ such that $m_r(\trans(\inst_1, \ldots, \inst_k, \bar{\alpha})) = h$ and $m_r(\trans(\inst_1, \ldots, \inst_k, \bar{\alpha}+\varepsilon)) = l$, where $h,l\in \labels$ and $l\neq h$.
%
Interestingly, given $\inst_1, \ldots, \inst_k$, Receiver can efficiently compute $\bar{\alpha}$ through the binary search procedure sketched in Algorithm~\ref{alg:binary}.

\begin{algorithm}[t]
\SetAlgoLined
\KwIn{$\inst_1, \ldots, \inst_k$; $\varepsilon > 0$}
$\alpha_{\textnormal{high}}$ := H\; 
 $\alpha_{\textnormal{low}}$ := L\;
 \Repeat{$\alpha_{\textnormal{low}} - \alpha_{\textnormal{high}} < \varepsilon$}{
  $\boldsymbol{x}_\textnormal{high}$ := $\trans$($\inst_1, \ldots, \inst_k, \alpha_{\textnormal{high}}$)\; 
  $\hat{y}_{\textnormal{high}}$ := $m_r(\inst_{\textnormal{high}})$\;
  $\boldsymbol{x}_{\textnormal{low}}$ := $\trans$($\inst_1, \ldots, \inst_k, \alpha_{\textnormal{low}}$)\; 
  $\hat{y}_{\textnormal{low}}$ := $m_r(\inst_{\textnormal{low}})$\;
  \If{$\hat{y}_{\textnormal{high}} = \hat{y}_{\textnormal{low}}$}{ \kwFailure{cannot be equal}}
  $\alpha_\textnormal{mid}$ := $(\alpha_{\textnormal{low}} + \alpha_{\textnormal{high}})/2$\;
  $\inst_{\textnormal{mid}}$ := $\trans$($\inst_1, \ldots, \inst_k, \alpha_{\textnormal{mid}}$)\; 
  $\hat{y}_{\textnormal{mid}}$ := $m_r(\inst_{\textnormal{mid}})$\;
  \eIf{$\hat{y}_{\textnormal{high}} \neq \hat{y}_{\textnormal{mid}}$}{
    $\alpha_{\textnormal{low}}$ := $\alpha_{\textnormal{mid}}$\;
  }{
   $\alpha_{\textnormal{high}}$ := $\alpha_{\textnormal{mid}}$\;
   }
 }
 \KwOut{$\widetilde{\inst} := \boldsymbol{x}_{\textnormal{high}}$, labels $h := \hat{y}_{\textnormal{high}}$ and $l := \hat{y}_{\textnormal{low}}$}
 \caption{Edge example binary search algorithm.}
 \label{alg:binary}
\end{algorithm}

The search procedure starts from the predefined interval [H, L].\footnote{In our experiments we use H = $0$ and L = $1/2$.}
At each iteration, two examples, i.e., $\inst_{\textnormal{high}}$ and $\inst_{\textnormal{low}}$, are generated and classified with Receiver's model $m_r$, so obtaining $\hat{y}_{\textnormal{high}}$ and $\hat{y}_{\textnormal{low}}$, respectively.
If $\hat{y}_\textnormal{high} = \hat{y}_\textnormal{low}$, the algorithm terminates with a failure and no edge example is returned.
Otherwise, the current search interval is split in half by computing $\alpha_{\textnormal{mid}}$, $\inst_\textnormal{mid}$, and $\hat{y}_\textnormal{mid}$.
Then, if $\hat{y}_\textnormal{high} \neq \hat{y}_\textnormal{mid}$, the algorithm iterates on the first half of the current interval.
Otherwise, if $\hat{y}_\textnormal{low} \neq \hat{y}_\textnormal{mid}$, the search procedure continues on the second half.
The loop terminates when the interval width goes under a threshold $\varepsilon$, which represents the granularity of a single position in the feature vectors of the example, e.g., a single pixel in the case of Figure~\ref{fig:sample-craft}.
Eventually, the algorithm returns the edge example $\inst_\textnormal{high}$, as well as the two class labels $h$ and $l$ associated with the last, smallest interval.

%
%
%
%
%
%


It is worth noticing that, since it does not explore the entire feature space, Algorithm~\ref{alg:binary} might fail to generate an edge examples on some inputs.
In general, the effectiveness of this heuristics depends on the choice of $\psi$.

At the end of the calibration phase, channel parameters generated by Receiver amount to the tuple $(f,\boldsymbol{\widetilde{x}},h,l)$.

\subsection{Bit transmission}
\label{SEC:BIASED_ID}

The transmission of one bit is based on the variations, during $f$ FL rounds, of $m(\widetilde{\inst})$ between $h$ and $l$.
In particular, by poisoning its local model, Sender drives $m(\widetilde{\inst})$ to assume the desired value, while Receiver monitors it to read the transmitted bit.
Transmissions are organized in consecutive frames of size $f$. 
Sender and Receiver are synchronized through an FL round counter $r \in \{1, \ldots, f\}$.
Sender follows the procedure given in Algorithm~\ref{alg:sender}.

\begin{algorithm}[t]
\SetAlgoLined
\KwIn{$f$, $\boldsymbol{\widetilde{x}}$, $h$, $l$}
 
 \Repeat{transmission completed}{
  \If{$r = 1$}{
    $b$ := nextBit()\;
    $v$ := $m(\boldsymbol{\widetilde{x}})$\;
    $v_\lnot$ := \leIf{$v = h$}{$l$}{$h$}
  }
  
  \If{selected by server for training}{
     $v_r$ := $m(\boldsymbol{\widetilde{x}})$\;
     \Switch{b}{
       \Case{0}{
         \lIf{$v_r \neq v$ }{train($m_s$, $\boldsymbol{\widetilde{x}}$, $v$)}
       }
       \Case{1}{
         \lIf{$v_r \neq v_\lnot$ }{train($m_s$, $\boldsymbol{\widetilde{x}}$, $v_\lnot$)}
       }
     }
    upload($m_s$)\;
  }
  $r$ := $(r \% f) + 1$ \;
 }
 \caption{Sender transmission algorithm.}
 \label{alg:sender}
\end{algorithm}


%
%
%
At the beginning of each transmission frame, i.e., when $r = 1$, Sender sets the next bit to be sent $b$, and uses the last received global model $m_r = m$ to classify $\boldsymbol{\widetilde{x}}$, thus obtaining $v \in \{h,l\}$. 
Also, Sender sets $v_{\lnot} \in \{h,l\}$ so that $v \neq v_{\lnot}$. 
At each round $r$, if Sender is selected by the FL server, it classifies $\boldsymbol{\widetilde{x}}$ with the global model $m$, so obtaining $v_r$. 
Then, depending on the bit to be sent, Sender trains its local model $m_s$ according to Table~\ref{tab:poison}. 

\begin{table}[t]
\caption{Sender's model poisoning cases.}
\label{tab:poison}
\begin{small}
\begin{center}
\begin{tabular}{c c c c}
     \bottomrule
     \multicolumn{2}{c}{\cellcolor{gray!25} $\boldsymbol{b = 0}$} & \multicolumn{2}{c}{\cellcolor{gray!25} $\boldsymbol{b = 1}$} \\
     \cellcolor{gray!25} $\boldsymbol{v_r = v}$ & \cellcolor{gray!25} $\boldsymbol{v_r = v_{\lnot}}$ & \cellcolor{gray!25} $\boldsymbol{v_r = v}$ & \cellcolor{gray!25} $\boldsymbol{v_r = v_{\lnot}}$ \\
     \toprule
     do nothing & train($m_s, \widetilde{\inst}, v$) & train($m_s, \widetilde{\inst}, v_{\lnot}$) & do nothing \\
     \toprule
\end{tabular}
\end{center}
\end{small}
\end{table}

Intuitively, the purpose of the operation above is to keep the label assigned to $\widetilde{\inst}$ by the global model when sending 0, and to flip it when sending 1.
This channel implementation amounts to a Differential Manchester encoding~\cite{Proakis2007digital}.
Eventually, Sender uploads its local model to the FL server.

\begin{algorithm}[t]
\SetAlgoLined
\KwIn{$f$, $\boldsymbol{\widetilde{x}}$, $h$, $l$}
 
 \Repeat{transmission completed}{
  \uIf{$r = 1$}{
    $v_1$ := $m(\boldsymbol{\widetilde{x}})$\;
  }
  \ElseIf{$r = f$}{
     $v_f$ := $m(\boldsymbol{\widetilde{x}})$ \;
     $b$ := \leIf{$v_1 = v_f$}{0}{1}
     received($b$)\;
  }
  $r$ := $(r \% f) + 1$ \;
 }
 \caption{Receiver bit test algorithm.}
 \label{alg:receiver}
\end{algorithm}

Concurrently, Receiver executes Algorithm~\ref{alg:receiver}.
Receiver uses the global model $m$ to classify $\boldsymbol{\widetilde{x}}$ both when $r=1$ and $r=f$, so obtaining $v_1$ and $v_{f}$, respectively.
Finally, Receiver reads 0 if $v_1 = v_{f}$ and 1 otherwise.

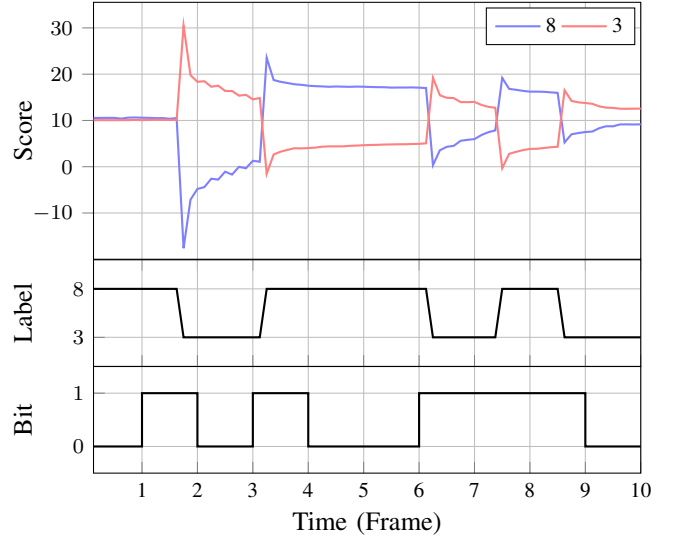
\begin{figure}[t]
\begin{tikzpicture}
\begin{axis}[ymin=-20, xmin=1, xmax=80, xlabel={}, ylabel={Score}, x label style={at={(axis description cs:0.5,-0.1)},anchor=center},
    y label style={at={(axis description cs:0.04,0.5)}, anchor=center},
    set layers=standard,
    xmajorgrids=true,
    ymajorgrids=true,
    ytick align=outside,
    ytick={-10,0,10,20,30},
    xtick={0,8,...,80},
    xticklabels={},
    xtick pos=left,
    ytick pos=left,
    yticklabel style={font=\footnotesize},
    xticklabel style={font=\footnotesize},
    height=5cm,
    width=\columnwidth,
    legend style={at={(0.99,0.98)},anchor=north east},
    legend columns=-1,
    ybar=0pt, bar width=0.1, bar shift=0pt,
    legend style={nodes={scale=0.8, transform shape}}
]
\addplot[line legend,draw=blue!50, sharp plot, thick] table [col sep=comma] {data/pattern9.csv}; 
\addlegendentry{8};
\addplot[line legend,draw=red!50, sharp plot, thick] table [col sep=comma] {data/pattern7.csv};
\addlegendentry{3};
\end{axis}

\begin{axis}[ymin=0, ymax=11, xmin=1, xmax=80, xlabel={}, ylabel={Label}, x label style={at={(axis description cs:0.5,-0.1)},anchor=center},
    y label style={at={(axis description cs:0.04,0.5)}, anchor=center},
    set layers=standard,
    xmajorgrids=true,
    ymajorgrids=true,
    ytick align=outside,
    ytick align=inside,
    xtick={0,8,...,80},
    xticklabels={},
    ytick={3,8},
    xtick pos=left,
    ytick pos=left,
    yticklabel style={font=\footnotesize},
    xticklabel style={font=\footnotesize},
    height=3cm,
    width=\columnwidth,
    xticklabels={,,},
    legend style={nodes={scale=0.8, transform shape}},
    at={(0,0)}, anchor=above north west
]
\addplot[draw=black,sharp plot,thick] table [col sep=comma] {data/patternmax.csv}; 
\end{axis}

\begin{axis}[ymin=-0.5, ymax=1.5, xmin=1, xmax=80, xlabel={Time (Frame)}, ylabel={Bit}, x label style={at={(axis description cs:0.5,-0.1)},anchor=center},
    y label style={at={(axis description cs:0.04,0.5)}, anchor=center},
    set layers=standard,
    xmajorgrids=true,
    ymajorgrids=true,
    ytick align=outside,
    ytick align=inside,
    xtick={0,8,...,80},
    xticklabels={,1,2,3,4,5,6,7,8,9,10},
    ytick={0,1},
    xtick pos=left,
    ytick pos=left,
    yticklabel style={font=\footnotesize},
    xticklabel style={font=\footnotesize},
    height=3cm,
    width=\columnwidth,
    legend style={nodes={scale=0.8, transform shape}},
    at={(0cm,-1.42cm)}, anchor=above north west
]
\addplot[color=black,thick] coordinates {
            (0, 0)
            (8, 0)
            (8, 1)
            (16, 1)
            (16, 0)
            (24, 0)
            (24, 1)
            (32, 1)
            (32, 0)
            (40, 0)
            (48, 0)
            (48, 1)
            (56, 1)
            (64, 1)
            (72, 1)
            (72, 0)
            (80,0)
        };
\end{axis}

\end{tikzpicture}
\caption{Transmission of bits 0101001110 over a covert channel.}
\label{fig:difman}
\end{figure}

In Figure~\ref{fig:difman}, we show as an example the transmission of 10 bits for the MNIST scenario discussed above, using $\widetilde{\inst}$ of Figure~\ref{fig:sample-craft}, $h=8$, and $l=3$.
The top diagram shows the internal scores assigned by the global model to the edge example $\boldsymbol{\widetilde{x}}$ during the transmission.
By classifying $\boldsymbol{\widetilde{x}}$, Receiver observes the alternation of labels 8 and 3 (center).
Then, Receiver interprets it as the sequence of bits 0101001110.

%

\section{Channel description}
\label{SEC:CHARAT}

For each edge example, our implementation provides a digital, broadcast channel supporting \emph{half duplex}\footnote{Intuitively, Sender and Receiver cannot transmit at the same time, but they can alternate their roles.} transmissions.
In principle, an attacker can even create several, parallel channels of this type to enlarge the communication bandwidth.
However, parallel channels may generate interference.
Below, we detail the features of a single channel.

In terms of capacity, we treat the covert channel as a standard BMC (see, Section~\ref{sec:bkg-covert}).
We just notice that the binary input of the channel is $m_b$, with $b \in \{0,1\}$.
Intuitively, inputs $m_b$ represent Sender's local models $m_s$ uploaded to the FL server.
More precisely, we use $m_b$ to distinguish between the two model poisoning cases used to transmit $b$ (see Table~\ref{tab:poison}).

In terms of communication quality, we consider BER and SNR, as discussed in Section~\ref{sec:bkg-covert}.
While BER is straightforward, defining SNR in our context requires more attention.
In general, SNR is computed by periodically sampling the channel signal $s$ and noise $n$ at the end of each transmission frame $t$.
Here, defining $s$ and $n$ is non-trivial, since our channel does not rely on a physical medium.
Indeed, a covert channel consists of two prediction labels, namely $h$ and $l$, and a perturbed example $\boldsymbol{\widetilde{x}}$.
For each label $i \in \{h, l\}$, at time $t$ we can measure the prediction score assigned by the global model to $i$ when classifying $\boldsymbol{\widetilde{x}}$ at each frame's end.
Also, assuming that prediction scores range within the interval $[Z, -Z]$,\footnote{$Z$ can be esteemed as the maximum score observed during a transmission.} we normalize the prediction scores by linearly scaling them in $[1,-1]$.
Thus, for each label $i$ the label signal $z_i(t)$ amounts to the normalized score described above. 
We define the overall received signal $z(t) = z_h(t) - z_l(t)$, i.e., as the differential, normalized signal.

Since a direct measure of the transmitted signal $s(t)$ cannot be computed, we approximate it to the differential signal that switches its intensity between $1$ and $-1$.
Intuitively, this is equivalent to stating that Sender attempts to transmit 0 by setting $z(t) \cdot z(t+1) = 1$ and 1 by setting $z(t) \cdot z(t+1) = -1$.\footnote{Notice that this assumption is more restrictive w.r.t. the actual implementation, since Receiver only requires $z(t) \cdot z(t+1) < 0$ and $z(t) \cdot z(t+1) > 0$ to read 0 and 1, respectively.}
Thus, for each frame $t$, we set 

\smallskip
\noindent
\begin{equation}
\label{EQ:signal_est}
    \begin{aligned}
    s(t) &=\: \frac{1 - 2b}{s(t-1)}, \\ 
    \end{aligned}
\end{equation}
\smallskip
\noindent
where $b$ is the bit transmitted during frame $t$, and, by construction, $s(0) = 1$.
Then, we define the noise at time $t$ as $n(t) = z(t) - s(t)$.

\begin{figure}[t]
    \begin{tikzpicture}
\begin{axis}[ymin=-1.1, ymax=1.5, xmin=1, xmax=80, xlabel={Time (FL rounds)}, ylabel={Intensity}, x label style={at={(axis description cs:0.5,-0.1)},anchor=center},
    y label style={at={(axis description cs:0.04,0.5)}, anchor=center},
    set layers=standard,
    xmajorgrids=true,
    ymajorgrids=true,
    ytick align=outside,
    ytick={-1,0,1},
    xtick={0,8,...,80},
    xticklabels={,1,2,3,4,5,6,7,8,9,10},
    xtick pos=left,
    ytick pos=left,
    yticklabel style={font=\footnotesize},
    xticklabel style={font=\footnotesize},
    height=5cm,
    width=\columnwidth,
    legend style={at={(0.99,0.98)},anchor=north east},
    legend columns=-1,
    ybar=0pt, bar width=0.1, bar shift=0pt,
    legend style={nodes={scale=0.8, transform shape}}
]
\addplot[line legend,draw=blue!50, sharp plot, dashed,thick] table [col sep=comma] {data/zhnorm.csv}; 
\addlegendentry{$z_h(t)$};
\addplot[line legend,draw=red!50, sharp plot, dashed,thick] table [col sep=comma] {data/zlnorm.csv};
\addlegendentry{$z_l(t)$};
\addplot[line legend,draw=purple, smooth, thick] table [col sep=comma] {data/znorm.csv};
\addlegendentry{$z(t)$};

\addplot[line legend, color=black!80,thick,sharp plot] coordinates {
            (0, 1)
            (8, 1)
            (16, -1)
            (24, -1)
            (32, 1)
            (40, 1)
            (48, 1)
            (56, -1)
            (64, 1)
            (72, -1)
            (80, -1)
        };
\addlegendentry{$s(t)$};
\end{axis}

    \node (B) at (2.05,1.1) {};
    \node (A) at (2.05,0.05) {};
    \draw[thick,latex-latex] (A) -- (B) node[midway,sloped,left,rotate=-90] {\footnotesize $n(3)$};

\end{tikzpicture}
    \caption{Signals and noise for the FL covert channel of Figure~\ref{fig:difman}.}
    \label{fig:SNR}
\end{figure}
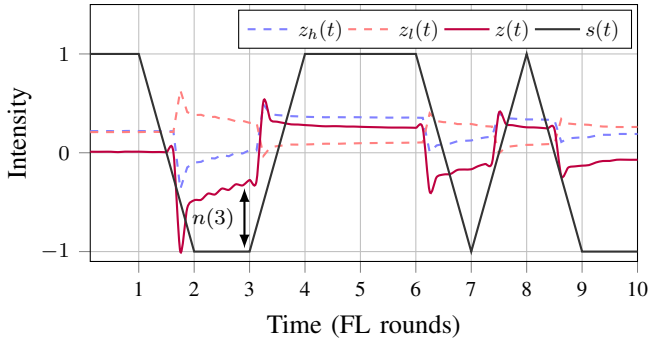

The overall intuition behind our definitions of $z(t)$, $s(t)$ and $n(t)$ is given in Figure~\ref{fig:SNR}. 
There, red and blue dashed lines denote $z_h$ and $z_l$ signals, i.e., the normalized version of the scores of Figure~\ref{fig:difman}.
Instead, the purple line denotes $z(t)$ and the gray line denotes $s(t)$.
Also, the vertical arrow shows the value of $n(t)$ at $t=3$.
From this and~\eqref{EQ:signal_est}, we define SNR of the covert channel as
%
%
\begin{equation}
    SNR = \frac{z(t)^2}{\sigma_n^2},  \label{EQ:SNR}
\end{equation}
where $\sigma_n$ is the standard deviation of the \emph{normalized} noise $\bar{n}(t)$ defined as $\bar{n}(t) = \tfrac{n(t)}{\max_{t'}\vert n(t') \vert}$.
\smallskip


\section{Conclusion}
\label{sec:conclusion}

In this paper, we introduced a new covert channel leveraging on FL systems.
Our attack allows a malicious agent to establish stealth communications between FL clients that should rather stay isolated.
This attack scenario paves the way for several exploits that we plan to investigate in the future.
Among them, we are particularly interested in attacks breaking the air gap.
These attacks appear realistic since many FL clients feed the training process with sensor-generated examples.
Other future directions include $(i)$ measuring the performance of the covert channel, $(ii)$ evaluating whether the channel can be detected, and $(iii)$ investigating possible countermeasures.




\vspace{1em}
\bibliographystyle{IEEEtran}
\bibliography{bibliography}

\end{document}